\documentclass[12pt,dvips]{article}

\usepackage{rotating}
\usepackage{hhline}
\usepackage{array}
\usepackage{epsfig}

\begin{document}

\renewcommand{\thefootnote}{\fnsymbol{footnote}}
\tolerance=100000

\newcommand{\lsim}{\raisebox{-0.13cm}{~\shortstack{$<$\\[-0.07cm] $\sim$}}~}
\newcommand{\gsim}{\raisebox{-0.13cm}{~\shortstack{$>$\\[-0.07cm] $\sim$}}~}

\newcommand{\imag}{\Im {\rm m}}
\newcommand{\real}{\Re {\rm e}}
\newcommand{\s}{\\ \vspace*{-3.5mm}}

\def\tablename{\bf Table}%
\def\figurename{\bf Figure}%

\begin{flushright}
DESY 06--011 \\[-0.1cm]
KAIST--TH 2006/02 \\[-0.1cm]
hep-ph/0602109\\
\today
\end{flushright}

\vspace{0.3cm}

\begin{center}
  {\large \bf Heavy Higgs Resonances for the Neutralino Relic Density\\[-1mm]
              in the Higgs Decoupling Limit of the CP--noninvariant\\
              Minimal Supersymmetric Standard Model
              }\\[1.2cm]
S.Y. Choi$^1$\footnote{E-mail: \texttt{sychoi@chonbuk.ac.kr}}
and Y.G. Kim$^2$\footnote{E-mail: \texttt{ygkim@muon.kaist.ac.kr}}
\end{center}

\vskip 0.2cm

{\small
\begin{center}
$^1$ {\it Deutches Elektronen--Synchrotron DESY, D--22603 Hamburg, Germany and \\
          Physics Department and RIPC, Chonbuk National University, Jeonju
          561-756, Korea}\\[2mm]
$^2$ {\it Department of Physics, KAIST, Daejon 305-701, Korea}
\end{center}
}

\renewcommand{\thefootnote}{\fnsymbol{footnote}}
\vspace{2.cm}

\begin{abstract}
\noindent
The lightest neutralino is a compelling candidate to account for cold dark matter
in the universe in supersymmetric theories with $R$--parity. In the CP--invariant
theory, the neutralino relic density can be found in accord with recent WMAP data
if neutralino annihilation in the early universe occurs via the $s$--channel $A$
funnel. In contrast, in the CP--noninvariant theory two heavy neutral Higgs bosons
can contribute to the Higgs funnel mechanism significantly due to a CP--violating
{\it complex} mixing between two heavy states, in particular, when they
are almost degenerate. With a simple analytic and numerical
analysis, we demonstrate that the CP--violating Higgs mixing can modify the
profile of the neutralino relic density {\it considerably} in the heavy Higgs
funnel with the neutralino mass close to half of the heavy Higgs masses.
\end{abstract}
%

%============================================================================
\newpage

The nature of the dark matter is one of the most important questions  at the
interface of particle physics and cosmology. Recently there have been
big improvements in the astrophysical and cosmological data, most notably due
to the Wilkinson microwave anisotropy probe (WMAP)\cite{WMAP} and the Sloan
digital sky survey (SDSS)\cite{SDSS}. With the data we can infer the following
$2\sigma$ range for the density of cold dark matter normalized by the critical
density
\begin{eqnarray}
0.094 < \Omega_{\rm CDM} h^2 < 0.129,
\label{eq:Omega_CDM}
\end{eqnarray}
where $h\approx 0.7$ is the (scaled) Hubble constant in units of
100 km/sec/Mpc. Such a precise determination of $\Omega_{\rm CDM} h^2$ imposes
severe constraints on any model that tries to explain it.\s

In supersymmetric theories with $R$--parity \cite{MSSM}, the lightest
supersymmetric particle (LSP), which is typically the lightest neutralino
$\tilde{\chi}^0_1 \equiv\chi$, is stable and it serves as an excellent
cold dark matter (CDM) candidate \cite{LSP_DM,SUSY_DM_review}.
However, typical mSUGRA models in the parameter space of minimal SUSY predict
much larger values for the neutralino relic density than the values in the
range (\ref{eq:Omega_CDM}). Some specific mechanisms leading to strongly
enhanced neutralino annihilation are required to produce the observed dark
matter relic density \cite{enhancement_mechanism}. Such an enhancement might be
due to the presence of light sleptons, enhancing the LSP annihilation into
leptons, to an accidental degeneracy of the LSP and the lighter stau (or stop),
leading to enhanced LSP--stau (or stop) co--annihilation, to the LSP with
significantly mixed gaugino--higgsino components, enhancing the annihilation
into gauge bosons, or to an accidental degeneracy $M_A\approx 2 m_\chi$ with
large $\tan\beta$, leading to enhanced annihilation through an $s$--channel
pseudoscalar $A$ in the CP--invariant theory.\s

In particular, the enhanced LSP annihilation via a $A$ funnel in the
CP--invariant case is due to two reasons; (i) the $S$--wave amplitude for
$\chi\chi\to A$ is not suppressed near threshold while the $P$--wave amplitude
for $\chi\chi\to H$ is suppressed near threshold and (ii) the total $A$ decay
width\footnote{The decays, $A \to W^+W^-$ and $A\to ZZ$, are forbidden, leading to
a small $A$ width for small $\tan\beta$ (unless the decay $A\to t\bar{t}$ is open).}
becomes large as the $A\to b\bar{b}$ decay mode is greatly enhanced for
large $\tan\beta$.\s

The generic feature of the $A$ funnel enhancement could, however, be greatly
modified due to the CP--violating mixing among neutral Higgs bosons
as well as due to the CP--violating Higgs couplings to neutralino pairs in
the CP--noninvariant theory \cite{cp_violating_higgs, pilaftsis_wagner,
complex_mixing_1, CKLZ, cp_noninvariant}. In this work we analyze, both
analytically and numerically, the impact on the LSP relic density by the
CP--violating Higgs mixing, loop--induced at the loop level in the
CP--noninvariant MSSM \cite{DM_MSSM_CP}. To be specific, we consider the
case when two (almost) degenerate heavy neutral Higgs bosons $H$ and $A$ are
essentially decoupled from the lightest neutral Higgs boson\footnote{This
situation is naturally realized in the MSSM  in the decoupling
limit with $M_A > 2 m_Z$ \cite{CKLZ, cp_invariant_decoupling}.} and
their masses are very close to twice the LSP mass.\s

With the lightest neutral Higgs boson decoupled, the CP--violating mixing of the
two nearly--degenerate heavy Higgs bosons is described by a $2\times 2$
{\it complex} mass matrix, composed of a real
dissipative part and an imaginary absorptive part \cite{CKLZ}. This mixing
can be very large, generating frequent mutual transitions inducing large
CP--odd mixing effects, which are quantitatively described by the complex mixing
parameter $X$:
\begin{eqnarray}
X = \frac{1}{2}\tan 2\theta
  = \frac{\Delta^2_{HA}}{M^2_H-M^2_A -i\left[M_H\Gamma_H-M_A\Gamma_A\right]},
\end{eqnarray}
where the complex off--diagonal term  $\Delta^2_{HA}$ of the
Higgs mass matrix couples two Higgs states.\s

The Higgs masses and widths are then shifted in a characteristic pattern by
the CP--violating mixing \cite{resonant_mixing_zerwas}, of which
the individual shifts can be obtained by separating real and imaginary parts
in the relations:
\begin{eqnarray}
 && \left[M^2_{H_2}\!-\!i M_{H_2} \Gamma_{H_2}\right]
  \!-\!\left[M^2_H\!-\!i M_H \Gamma_H\right]
   =
  - \left\{\left[M^2_{H_3}\!-\!i M_{H_3} \Gamma_{H_3}\right]
  \!-\!\left[M^2_A\!-\!i M_A \Gamma_A\right]\right\}\nonumber\\[2mm]
 &&{ }\hskip 2.32cm = -\left\{\left[M^2_A\!-\!i M_A \Gamma_A\right]
         \!-\!\left[M^2_H\!-\!i M_H \Gamma_H\right]\right\}
     \times\,\mbox{\small $\frac{1}{2}$}\, [\sqrt{1+4X^2}-1]
\end{eqnarray}
In such a non--Hermitian mixing the ket and bra mass eigenstates have
to be defined separately: $|H_i \rangle = C_{i\alpha}|H_\alpha\rangle$
and $\langle\widetilde{H}_i| = C_{i\alpha}\langle H_\alpha|$ ($i=2, 3$ and
$H_\alpha=H,A$); $C_{2H}=\cos\theta, C_{2A}=\sin\theta, C_{3H}=-\sin\theta$
and $C_{3A}=\cos\theta$ in terms of the complex mixing angle $\theta$.\s

As two mass eigenstates have no definite CP parity and an enlarged mass
splitting, the profile of the LSP relic density can considerably be modified
in the heavy Higgs funnel. For a simple analytic and numerical illustration,
we consider a specific scenario within the CP--violating MSSM [MSSM--CP], while
a more comprehensive analysis is separately given in a future publication.
We assume the source of CP violation to be localized entirely in the complex stop
trilinear coupling $A_t$ but all the other interactions to be CP
conserving.\footnote{This assignment is compatible with the bounds
from the electric dipole moment measurements \cite{EDM}.}\s

In this situation,  CP violation is transmitted through stop--loop corrections
to the effective Higgs potential, generating three CP--odd complex quartic
parameters. The effective parameters have been calculated in
Ref.~\cite{pilaftsis_wagner} to two--loop accuracy and, with $t/\tilde{t}$
contributions, the parameters are determined by the parameters; the SUSY scale
$M_S$ which is taken to be essentially the average of two stop masses--squared,
the higgsino parameter $\mu$, the stop trilinear parameter $A_t$ and the
top Yukawa coupling $h_t=\sqrt{2} \bar{m}_t /v\sin\beta$ defined with the running
$\overline{\rm MS}$ mass $\bar{m}_t$ and the Higgs vacuum expectation value
$v\approx 246$ GeV. The one--loop improved Born Higgs mass matrix is derived
from this effective Higgs potential and then the matrix elements are shifted to
the pole--mass parameters by including dispersive contributions from Higgs
self--energies.\s

Before evaluating the impact of the complex $H/A$ mixing on the
LSP relic density in the heavy--Higgs funnel, we describe an approximate
procedure for estimating the relic density \cite{Griest_Seckel}. The LSP
number density  is evolved in time
according to the Boltzmann equation. When the temperature of the Universe is
higher than the LSP mass, the number density is simply given by its
thermal--equilibrium density. However, once the temperature drops below
the LSP mass, the number density drops exponentially. As a result, the LSP
annihilation rate becomes smaller than the Hubble expansion rate at a certain
point when the LSP neutralinos fall out of equilibrium and the LSP number
density in a co--moving volume remains constant. The present LSP relic
abundance is then approximately given by
\begin{eqnarray}
\Omega h^2 \simeq {1.07 \times 10^9~ \rm GeV^{-1} \over J g_*^{1/2}\, M_{PL}},
\label{eq:omega}
\end{eqnarray}
where $g_*=81$ is the number of relativistic degrees of freedom
and $\rm M_{PL}=1.22 \times 10^{19}\, GeV$ is the Planck mass.
And the integral $J$ is given by
\begin{eqnarray}
J(x_f) = \int_{x_f}^{\infty} {\langle\sigma v\rangle \over x^2} dx,
\label{eq:j_xf}
\end{eqnarray}
where $\langle \sigma v\rangle$ is the thermally averaged LSP annihilation
cross section times the relative velocity $v$ of two annihilating LSPs, and
$x_f=m_\chi/T_f$ with the freeze--out temperature $T_f \simeq m_\chi/25$ for
typical weak--scale numbers. We take $x_f=25$ in the following numerical
demonstration.\s

When the heavy Higgs boson masses are large and close to twice the LSP mass,
the LSP annihilation is dominated by heavy Higgs--boson exchanges.
The LSP annihilation rate can then be estimated with reasonable approximation by
including only the $s$--channel heavy Higgs boson exchanges.
In the decoupling limit the $H\chi\chi$ and $A\chi\chi$ couplings read
\begin{eqnarray}
&& \langle \chi_L|H|\chi_R\rangle = \langle \chi_R|H|\chi_L\rangle^*
  \simeq -\frac{g}{2}\,\, (N_{12}-N_{11}\tan\theta_W)
     (\sin\beta N_{13}+\cos\beta N_{14}), \nonumber\\
&& \langle \chi_L|A|\chi_R\rangle\, = \langle \chi_R|A|\chi_L\rangle^*\,
 = -\frac{g}{2}\, i(N_{12}-N_{11}\tan\theta_W)
     (\sin\beta N_{13}-\cos\beta N_{14}),
\label{eq:coupling}
\end{eqnarray}
in terms of $\tan\beta$ and the neutralino mixing matrix $N_{i\alpha}$
($i,\alpha=1$-$4$) diagonalizing the neutralino mass matrix ${\cal M}_N$ as
$N^*{\cal M}_N N^\dagger = {\cal M}_{diag}$ \cite{neutralino_mixing}.
The LSP annihilation rate multiplied by the relative velocity $v$ of two LSPs
can be expressed as
\begin{eqnarray}
\sigma v = \frac{1}{2} \sum_{a,b = H, A} {\cal P}_a {\cal P}^*_b\,\,
           \frac{\Gamma_{ab}(\sqrt{s})}{\sqrt{s}},
\label{eq:event_rate}
\end{eqnarray}
where the relative velocity $v$ is taken to be $2\beta=2\sqrt{1-4m^2_\chi/s}$,
and the production amplitudes ${\cal P}_{a,b}$ and the transition decay widths
$\Gamma_{ab}$ are defined as
\begin{eqnarray}
&& {\cal P}_a = \sum_{i=2,3}\, \sum_{b=H,A} C_{ia}\, \Pi_i\, C_{ib} \,
                 P(\chi\chi\to a),\nonumber\\
&& \Gamma_{ab} = \frac{1}{2\sqrt{s}}\sum_F \oint d\Phi_F\,
                 D(a\to F) D^*(b\to F),
\end{eqnarray}
with the Higgs propagators $\Pi_i=1/(s-M^2_{H_i}+i M_i\Gamma_{H_i})$. Here,
$P(\chi\chi\to H, A)$ are the $\chi\chi\to H, A$ production amplitudes,
determined by the couplings (\ref{eq:coupling}), and $D(H, A\to F)$
the $H, A\to F$ decay amplitudes, for any kinematically and dynamically
allowed decay mode $F$.
Evaluating $J(x_f)$ in Eq.~(\ref{eq:j_xf}) with the event rate
(\ref{eq:event_rate}) and inserting its value into
Eq.~(\ref{eq:omega}) yields the present neutralino relic density.\s

Although it is possible to calculate the masses and (transition) decay widths
of the heavy Higgs bosons fully, we estimate them in the present work
with a few approximations, which are reliable in the Higgs decoupling
limit. In general, the light Higgs boson, the fermions and electroweak gauge
bosons, and in supersymmetric theories, gauginos, higgsinos and scalar states
may contribute to the loops in the complex mass matrix. In the decoupling
limit, the couplings of the heavy Higgs bosons to gauge bosons and their
superpartners are suppressed. Assuming all the other supersymmetric particles
to be suppressed either by couplings or by phase space, we consider only
loops by the LSP neutralino, the light Higgs boson and the top/bottom quark
for the absorptive parts as characteristic examples; loops from other
(s)particles could be treated in the same way of course.\s

In order to demonstrate the effect of the CP--violating $H/A$ mixing
on the neutralino relic density in the MSSM--CP numerically, we adopt a
typical set of parameters\footnote{Analyses of electric dipole moments show that
the phase of $\mu$ is quite small, unless sfermions are very heavy \cite{EDM};
therefore its phase is set zero in our numerical demonstration.},
\begin{eqnarray}
M_S = 0.5\,\, {\rm TeV},\quad |A_t|= 1.0\,\, {\rm TeV}, \quad
\mu= 2.0\,\, {\rm TeV}; \quad \tan\beta=5,
\label{eq:parameter_set}
\end{eqnarray}
while varying the pseudoscalar mass $M_A$, the SU(2) gaugino mass $M_2$, and
the phase $\Phi_A$ of the trilinear term $A_t$, and taking $M_1\simeq  0.5 M_2$.
[By reparameterization of the fields, $M_2$ is set real and positive.]
For such a large $\mu$ compared to $M_2$, the LSP is almost
bino--like and its mass is close to $M_1$.\s

\begin{figure}[htb]
\begin{center}
\mbox{ }\\[0.2cm]
\epsfig{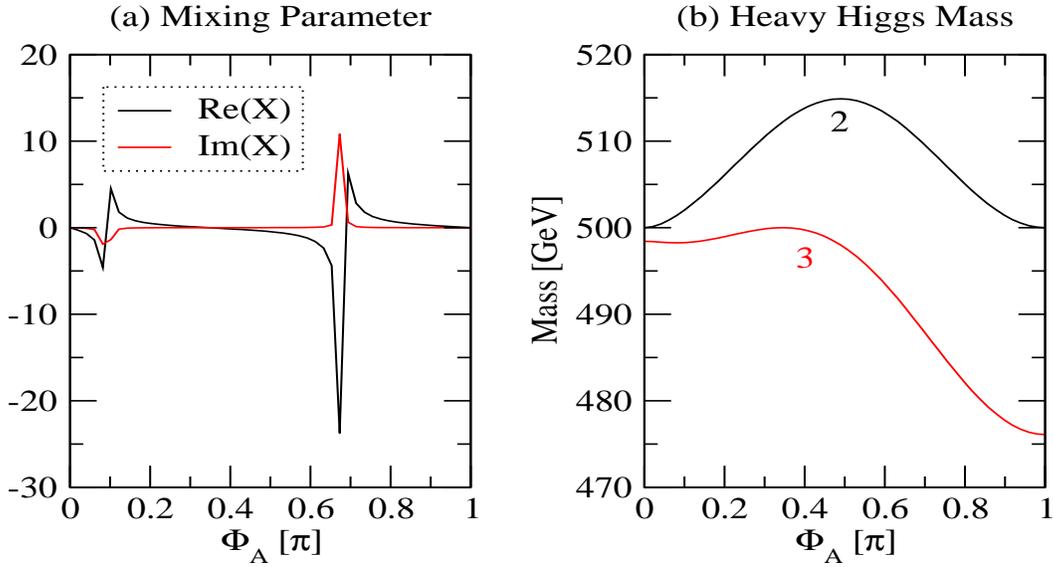}\hskip 0.5cm
\caption{\it The $\Phi_A$ dependence of (a) the real (black) and imaginary
            (red) parts of the mixing parameter $X$ and (b) the heavy Higgs
            boson masses, $M_{H_2}$ (black) and $M_{H_3}$ (red).
            $M_2$ and $M_A$ are set to 500 GeV.
            Note that $\real/\imag X(2\pi-\Phi_A) =+\real/-\imag X(\Phi_A)$ and
            the masses and widths are symmetric about $\Phi_A=\pi$.}
\label{fig:mixing}
\end{center}
\end{figure}

The $\Phi_A$ dependence of the $H/A$ mixing parameter $X$ and the heavy
Higgs masses and are displayed in Figs.~\ref{fig:mixing}(a) and (b),
respectively, for $M_{2,A}=0.5$ TeV.\footnote{With one common phase $\Phi_A$, the
complex mixing parameter $X$ obeys the relation $X(2\pi-\Phi_A)=X(\Phi_A)$ so that
all CP--even quantities are symmetric when switching from $\Phi_A$ to
$2\pi-\Phi_A$. Therefore we can restrict the discussion to the range $0\leq
\Phi_A\leq \pi$.} The two--state system in the MSSM--CP shows a very
sharp resonance CP--violating mixing, purely imaginary near $\Phi_A=0.09\pi$
and $\Phi_A=0.67\pi$. We note that the mass shift is indeed enhanced by more
than an order of magnitude if the CP--violating phase rises to non--zero values,
reaching a maximal value of the mass difference $\sim 24$ GeV.
As a result, the two mass eigenstates become clearly distinguishable,
incorporating significant admixtures of CP--even and CP--odd components
mutually in the wave functions.\s

\begin{figure}[htb]
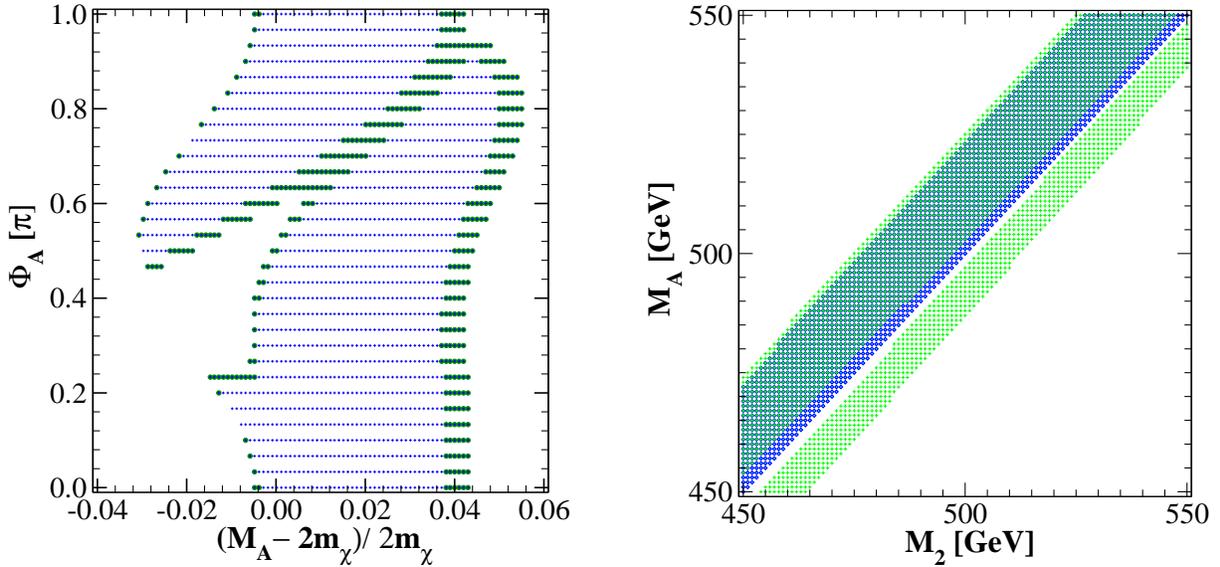

\begin{center}
\mbox{ }\\[0.3cm]
\epsfig{file=maphi_scan.eps,height=7.5cm,width=7.5cm}\hskip 1.cm
\epsfig{file=m2ma_scan.eps,height=7.5cm,width=7.5cm}
\vskip -0.1cm
\caption{\it Left panel: The allowed phase space of the CP phase $\Phi_A$ and
             the normalized mass difference $(M_A-2m_\chi)/2m_\chi$ for the
             range (\ref{eq:Omega_CDM}). The green  area is for the range
             (\ref{eq:Omega_CDM}), but the blue area is for the enlarged
             range with the lower bound ignored. Right panel: The allowed region
             of the $(M_2, M_A)$ plane for the bound $\Omega h^2 <0.129$
             in the CP--invariant case with $\Phi_A=0$ (a blue strip) and
             CP--noninvariant case with $\Phi_A=0.55\pi$ (two green strips).
             The values of the other relevant parameters are given in the
             text.}
\label{fig:fig2}
\end{center}
\end{figure}

The left panel of Fig.~\ref{fig:fig2} shows the allowed space of the phase
$\Phi_A$ and the normalized mass difference $(M_A-2m_\chi)/2m_\chi$ for the
range ($\ref{eq:Omega_CDM}$). Here we have set $M_2$ to 0.5 TeV and have
scanned the parameter space where $ 450\, {\rm GeV} \leq M_A \leq 550\, \rm GeV$
and $0 \leq \Phi_A \leq \pi$.
The allowed region for $\pi\leq \Phi_A \leq 2\pi$ is simply obtained by
reflecting the allowed region for $0\leq \Phi_A\leq \pi$ with respect to
$\Phi_A=\pi$. The green strip is for the range
(\ref{eq:Omega_CDM}) and the blue region for $\Omega h^2 < 0.095$.
In the other remaining region, we have $\Omega h^2 > 0.129$. One can clearly
see that (i) the neutralino relic density is indeed greatly suppressed for
$M_A \sim 2 m_\chi$ due to the Higgs resonances and the detailed
prediction for the relic density depends strongly on the value of the phase
$\Phi_A$ as well as the mass difference between $M_A$ and $2m_\chi$.\s

The right panel of Fig.~\ref{fig:fig2} shows the allowed regions of the
$(M_2, M_A)$ plane for $\Omega h^2 <0.129$ in the CP--invariant case
(one blue strip) with $\Phi_A=0$ and in the CP--noninvariant case with
$\Phi_A \simeq 0.55\pi$ (two green strips). Clearly, in order to satisfy
the relic density constraint, the LSP mass, which is approximately $0.5 M_2$,
should be close to half of the Higgs masses.
In the CP--invariant case only the CP--odd Higgs boson $A$ is active for
the Higgs funnel mechanism and so only one allowed strip with its width of
about 20 GeV is developed. In contrast, in the CP--noninvariant case
with $\Phi_A=0.55\pi$, both of the heavy Higgs bosons become active for
the funnel mechanism, leading to two strips; one strip is almost
identical to the strip in the CP--invariant case, but the other is newly
developed as the $H_3$ state, which is purely CP--even in the
CP--invariant case, has a significant CP--odd component due to
the CP--violating Higgs mixing. The combined width of two strips is widened
due to the enlarged mass splitting between two mass eigenstates in the
CP--noninvariant case.\footnote{Two Higgs--boson masses are approximately
$M_{H_2}\approx  M_A+ 15$ GeV and $M_{H_3}\approx M_A-5$ GeV, respectively.}\s

{\it To summarize.} We have examined the effect of the CP--violating H/A mixing
on the LSP annihilation cross section in the Higgs decoupling limit. By a simple
analysis with a specific parameter set (\ref{eq:parameter_set})
we have demonstrated that the CP--violating mixing can modify the profile of
the LSP relic density {\it considerably} in the heavy Higgs funnel with the
LSP mass close to half of the Higgs masses. Therefore, in order to
elucidate the Higgs funnel mechanism through high--energy experiments on the
supersymmetric particles, it is necessary to determine with good accuracy the complex
mixing angle between two Higgs states in addition to the LSP and heavy
Higgs boson masses and couplings \cite{heavy_Higgs_masses}.\s

\vskip 0.3cm

%-------------------------------
\subsection*{Acknowledgments}
%-------------------------------

The authors are grateful to P.M. Zerwas for his valuable comments and
suggestions. The work of SYC was supported partially by KOSEF through CHEP at
 Kyungpook National University and  by the Korea Research Foundation Grant
by the Korean Government (MOEHRD) (KRF--2006--013--C00097) and the work of
YGK was supported partially by the KRF Grant funded by the Korean
Government (KRF--2005--201--C00006) and  by the KOSEF Grant
(KOSEF R01--2005--000--10404--0).


\begin{thebibliography}{99}

\bibitem{WMAP} WMAP Collaboration, D.N. Spergel {\it et al.}, Astrophys. J.
   Suppl. Ser. {\bf 148} (2003) 175.

\bibitem{SDSS} SDSS Collaboration, M. Tegmark {\it et al.}, Phys. Rev.
   D {\bf 69} (2004) 103501.

\bibitem{MSSM} For a recent review, we refer to D.J.H. Chung, L.L. Everett,
   G.L. Kane, S.F. King, J.D. Lykken and L.T. Wang, Phys. Rept. {\bf 407}
   (2005) 1.

\bibitem{LSP_DM} H. Goldberg, Phys. Rev. Lett. {\bf 50} (1983) 1419; J. Ellis,
   J. Hagelin, D. Nanopoulos and M. Srednicki, Phys. Lett. B {\bf 127} (1983)
   233; J. Ellis, J. Hagelin, D. Nanopoulos, K. Olive and M. Srednicki,
   Nucl. Phys. B {\bf 238} (1984) 453.

\bibitem{SUSY_DM_review} For reviews on supersymmetric dark matter, see, for
   example, G. Jungman, M. Kamionkowski and K. Griest, Phys. Rept. {\bf 267}
   (1996) 195; G. Bertane, D. Hooper and J. Silk, Phys. Rept. {\bf 405} (2005)
   279.

\bibitem{enhancement_mechanism} H. Baer, C. Balazs, A. Belyaev, J.M. Mizukoshi,
   X. Tata and Y. Wang, JHEP {\bf 0207} (2002) 050; H. Baer and C. Balazs, JCAP
   {\bf 0305} (2003) 006; U. Chattopadhyay , A. Corsetti and P. Nath, Phys.
   Rev. D {\bf 68} (2003) 035005; J. Ellis, K.A. Olive, Y. Santoso and
   V.C. Spanos, Phys. Lett. B {\bf 565} (2003) 176; M. Battaglia, A. De Roeck,
   J.R. Ellis, F. Gianotti, K.A. Olive and L. Pape, Eur. Phys. J. C {\bf 33}
   (2004) 273; R. Arnowitt and B. Dutta and B. Hu, hep--ph/0310103; J.R. Ellis,
   K.A. Olive, Y. Santoso and V.C. Spanos, Phys. Rev. D {\bf 69} (2004) 095004;
   M.E. Gomez, T. Ibrahim, P. Nath and S. Skadhauge, Phys. Rev. D {\bf 70}
   (2004) 035014; J.R. Ellis, S. Heinemeyer, K.A. Olive and G. Weiglein,
   JHEP {\bf 0502} (2005) 013; A. Djouadi, M. Drees and J.L. Kneur,
   JHEP {\bf 0108} (2001) 055; C. Boehm, A. Djouadi and M. Drees,
   Phys. Rev. D {\bf 62} (2000) 035012; R. Arnowitt, B. Dutta and Y. Santoso,
   Nucl. Phys. B {\bf 606} (2001) 59; J.R. Ellis, K.A. Olive and Y. Santoso,
   Astropart. Phys. {\bf 18} (2003) 395; A. Djouadi, M. Drees and J.L. Kneur,
   Phys. Lett. B {\bf 624} (2005) 60.

\bibitem{cp_violating_higgs} A. Pilaftsis, Phys. Rev. D {\bf 58} (1998) 096010;
   Phys. Lett. B 435 (1998) 88; ; D.A. Demir, Phys. Rev. D {\bf 60} (1999)
   055006; S.Y. Choi, M. Drees and J.S. Lee, Phys. Lett. B {\bf 481} (2000) 57;
   M. Carena, J. Ellis, A. Pilaftsis and C.E.M. Wagner, Nucl. Phys. B {\bf 586}
   (2000) 92; Phys. Lett. B {\bf 495} (2000) 155; T. Ibrahim and P. Nath,
   Phys. Rev. D {\bf 63} (2001) 035009; S.W. Ham, S.K. Oh, E.J. Yoo, C.M. Kim
   and D. Son, Phys. Rev. D {\bf 68} (2003) 055003.

\bibitem{pilaftsis_wagner} A. Pilaftsis and C.E.M. Wagner, Nucl. Phys. B
   {\bf 553} (1999) 3.

\bibitem{complex_mixing_1} A. Pilaftsis, Nucl. Phys. B {\bf 504} (1997) 61;
   J.R. Ellis, J.S. Lee and A. Pilaftsis, Phys. Rev. D {\bf 70} (2004) 075010;
   J.S. Lee, hep-ph/0409020; J.R. Ellis, J.S. Lee and A. Pilaftsis, Nucl. Phys.
   B {\bf 718} (2005) 247; J.R. Ellis, J.S. Lee and A. Pilaftsis, Phys. Rev.
   D {\bf 71} (2005) 075007.

\bibitem{CKLZ} S.Y. Choi, J. Kalinowski, Y. Liao and P.M. Zerwas, Eur. Phys.
   J. C {\bf 40} (2005) 555.

\bibitem{cp_noninvariant} J.F. Gunion, H.E. Haber and J. Kalinowski,
   in preparation; For a different point of view see also I.F. Ginzburg,
   M. Krawczyk and P. Osland, hep--ph/0211371; M.N. Dubinin and A.V. Semenov,
   Eur. Phys. J. C {\bf 28} (2003) 233.

\bibitem{DM_MSSM_CP} P. Gondolo and K. Freese, JHEP {\bf 0207} (2002) 052;
    S.Y. Choi, S.-C. Park, J.H. Jang and H.S. Song, Phys. Rev. D {\bf 64}
    (2001) 015006; T. Nihei and M. Sagawa, Phys. Rev. D {\bf 70} (2004) 055011;
    T. Nihei, hep--ph/0508285; C. Balazs, M. Carena, A. Menon,
    D.E. Morrissey and C.E.M. Wagner, Phys. Rev. D {\bf 71} (2005) 075002.

\bibitem{cp_invariant_decoupling} J.F. Gunion and H.E. Haber, Phys. Rev. D
   {\bf 67} (2003) 075019.

\bibitem{resonant_mixing_zerwas} S. G\"{u}sken, J.H. K\"{u}hn and P.M. Zerwas,
    Nucl. Phys. {\bf B} 262 (1985) 393.

\bibitem{EDM} Y. Kizukuri and N. Oshimo, Phys. Lett. {\bf B249} (1990) 449;
    I. Ibrahim and P. Nath, Phys. Rev. D {\bf 58} (1998) 111301; {\bf 60}
    (1998) 099902(E); {\bf 61} (2000) 093004; M. Brhlik, G.J. Good and
    G.L. Kane, Phys. Rev. D {\bf 59} (1999) 115004; R. Arnowitt, B. Dutta and
    Y. Santos, Phys. Rev. D {\bf 64} (2000) 113010; V. Barger, T. Falk, T. Han,
    J. Jiang, T. Li and T. Plehn, Phys. Rev. D {\bf 64} (2001) 056007;
    S.Y. Choi, M. Drees and B. Gaissmaier, Phys. Rev. D {\bf 70} (2004) 014010.

\bibitem{Griest_Seckel} K. Griest and D. Seckel, Phys. Rev. D {\bf 43} (1991)
    3191.

\bibitem{neutralino_mixing} For details on the neutralino mixing in the MSSM,
   see, for instance, S.Y. Choi, J. Kalinowski, G. Moortgat--Pick and
   P.M. Zerwas, Eur. Phys. J. C {\bf 22} (2001) 563; {\bf 23} (2002) 769.

\bibitem{heavy_Higgs_masses} K. Desch, T. Klimkovich, T. Kuhl and A. Raspereza,
   hep--ph/0406229; M. Battaglia, hep-ph/0410123; M. Battaglia and M.E. Peskin,
   hep--ph/0509135.

\end{thebibliography}
\end{document}